\documentclass[12pt,preprint]{aastex}
\usepackage{graphicx}
\usepackage[CaptionAfterwards]{fltpage}
\slugcomment{Published in Nature on 14 October 2010.}

\shortauthors{Jewitt et al.}

\begin{document}

\title{Newly Disrupted Main Belt Asteroid \mbox{P/2010 A2}}

\author{David Jewitt$^{1,2,3}$, Harold Weaver$^4$, Jessica Agarwal$^5$, Max Mutchler$^6$, Micha{\l} Drahus$^1$}
\affil{$1$ Dept. Earth and Space Sciences, UCLA \\
$2$ Institute for Geophysics and Planetary Physics, UCLA \\
$3$ Dept. Physics and Astronomy, UCLA \\
$4$ The Johns Hopkins University Applied Physics Laboratory \\
$5$ ESA-ESTEC, Noordwijk, The Netherlands \\
$6$ Space Telescope Science Institute, Baltimore}
\email{jewitt@ucla.edu}

\textbf{Most main-belt asteroids are primitive rock and metal bodies
in orbit about the Sun between Mars and Jupiter. Disruption, through
high velocity collisions or rotational spin-up, is believed to be
the primary mechanism for the production and destruction of small
asteroids$^{1,2}$ and a contributor to dust in the Sun's Zodiacal
cloud$^3$, while analogous collisions around other stars feed dust
to their debris disks$^4$. Unfortunately, direct evidence about the
mechanism or rate of disruption is lacking, owing to the rarity of
events. Here we present observations of \mbox{P/2010 A2}, a
previously unknown inner-belt asteroid with a peculiar, comet-like
morphology that is most likely the evolving remnant of a recent
asteroidal disruption. High resolution Hubble Space Telescope
observations reveal an approximately 120 meter diameter nucleus with
an associated tail of millimeter-sized dust particles formed in
February/March 2009, all evolving slowly under the action of solar
radiation pressure.}


Object \mbox{P/2010 A2} was first detected on 2010 January 6 in data
from the US Airforce LINEAR survey telescope$^5$ and was immediately
classified as a short-period comet, based on the orbit and a diffuse
appearance presumably caused by ejected dust. The orbital elements,
however, are those of a main-belt asteroid (semimajor axis $a$ =
2.290 AU, eccentricity $e$ = 0.1244, inclination $i$ = 5.25$\degr$),
placing \mbox{P/2010 A2} in the newly recognized class of objects
known as main-belt comets$^6$.

Ground-based observations of \mbox{P/2010 A2} taken in early
January$^{7,8}$ revealed a peculiar morphology that was unlike any
previously observed comet, suggesting an origin other than by the
normal cometary process of water ice sublimation. The dust appeared
in a thin, parallel-sided tail (sometimes called a ``trail'')
detached from the nucleus, whereas typical cometary tails have their
origin in a dust coma surrounding the nucleus and are fan-shaped.
Hubble Space Telescope (HST) images taken at higher angular
resolution (Supplement Table~1, Figure~1) confirm a point-like
nucleus (N) at the leading edge of a thin, diffuse tail in which are
embedded crossed filamentary structures (AA and BB). The filaments
are the source of particles for the tail, including several dust
streaks barely resolved even at HST resolution (F). Several
persistent but faint and diffuse sub-nuclei (C) appear along the
filaments. The filament morphology did not change as the Earth
crossed the orbital plane of \mbox{P/2010 A2} on 2010 February~9
(e.g. compare images from Jan~29 and Feb~22 in Figure~2, at plane
angles -0.9$\degr$ and +0.9$\degr$, respectively), showing that the
filaments and sub-nuclei are not confined to the plane.

Unlike other main-belt comets$^6$, \mbox{P/2010 A2} orbits in the
inner regions of the belt where \mbox{S-type} asteroids are most
abundant$^9$. The \mbox{S-types} are refractory rocks, dominated by
materials formed at high temperatures, not by ice. Indeed, ice is
thermodynamically unstable at the 184~K temperature expected of an
isothermal blackbody at 2.29~AU. Primary nucleus, N, is unresolved
and fades in accordance with the inverse square law (See
Supplementary Information). These properties together show that the
nucleus is inert, with an estimated radius of $\sim$60~meters
(Supplementary Information).

\mbox{P/2010 A2} shows modest morphological evolution on timescales
of months (Figure~2), owing to changes in the distance (and
resolution), the observing perspective and intrinsic changes in the
object. Notable are a steady change in the position angle and a
narrowing of the tail (Table~3). From a model of dust motions
including solar gravity and radiation pressure$^{10}$, we calculate
the expected tail position angle as a function of the date of
emission of the dust particles from the nucleus. Particles emitted
at a given time with negligible relative speed lie on straight lines
emerging from the nucleus (``synchrones''), with larger particles
being closer to the nucleus. The position angles of these synchrones
are plotted in Figure~3, from which we infer dates of ejection in
2009 February/March, in agreement with another determination$^{11}$.
The dust dynamical model is a function of $\beta$, the ratio of the
radiation pressure force on a particle to the solar gravitational
attraction. We find that particles in the field of view of the HST
observations have $\beta < 2 \times$ 10$^{-4}$, corresponding to
particle sizes larger than 1~mm rising to $\sim$1~cm near the
crossed filaments. The narrowing of the tail (Figure~2) occurs
because particles launched perpendicular to the orbit reach maximum
height above the orbit plane one quarter orbit (10~months) after
ejection. The width of the dust tail implies out-of-plane dust
velocities $\delta v \sim$ 0.2 m~s$^{-1}$. Relative velocities
measured between the nucleus N and sub-nuclei in the filaments (e.g.
between N and C in Figure~1) are $\delta v <$ 0.2 m~s$^{-1}$.

The effective scattering cross-section of the dust tail is
comparable to the area of a circle of radius $r_e$ = 2100 m. If
contributed by particles in the mm to cm size range, this
cross-section corresponds to a dust mass $M$ = (6 to 60)
$\times$10$^7$ kg, equivalent to a sphere of the same density and
having a radius $r$ = 17 to 36~m (See Supplementary Information).

One possibility is that \mbox{P/2010 A2} was disrupted by rotational
bursting, perhaps caused by spin-up under the action of radiation
torques (the timescale for spin-up is very uncertain but it can be
$<$10$^5$ yr for a sub-kilometer body$^{12,13}$). If the dust
following \mbox{P/2010 A2} was produced by an impact, $r$ gives an
upper limit to the radius of the projectile, $r_p$ since, in a
hypervelocity impact, orders-of-magnitude more mass is ejected from
the target than is delivered by the projectile. We infer that the
projectile was of the order of a few meters in radius, tiny compared
to the primary nucleus. The velocity dispersion among asteroids in
the main belt is $\Delta V \sim$ 5 km~s$^{-1}$ ($^1$). From these
parameters we infer that the energy per unit target mass in the
responsible impact was $E/M$ = 1/2 [$r_p/r_n]^3 \Delta V^2 <$ 10$^5$
to 10$^6$ J~kg$^{-1}$, which encompasses the $E/M$ needed for
catastrophic fragmentation in a direct impact$^{14}$.
Experiments$^{15}$ and calculations$^{16}$ show that most mass in
hypervelocity impacts is displaced at low velocity, consistent with
the speeds measured.

The expected interval between collisional disruptions of 0.1~km
diameter asteroids in the main-belt is $\sim$1 yr$^{~17}$, while
damaging but non-disruptive impacts should be more frequent. The
$>$1 yr duration of visibility of the \mbox{P/2010 A2} debris cloud
suggests that we should expect to find one or more similar objects
at any time, in any all-sky survey with sensitivity equal to that of
LINEAR or greater. Comparable disruption events occurring annually
will release into the Zodiacal cloud about 2 to 20 kg~s$^{-1}$ of
dust, on average. This is only 0.1 to 1\% of the 600-1000~kg
s$^{-1}$ mass injection rate needed to keep the Zodiacal cloud in
steady state$^{18}$, suggesting that most of the mass comes from
comets$^{19}$ or another source.

\clearpage

\noindent \textbf{Supplementary Information} is linked to the online version of the paper at

www.nature.com/nature.

\noindent \textbf{Acknowledgements}

D.~J. thanks J.~Annis and M.~Soares-Santos for taking initial
observations at the WIYN telescope from which the unusual appearance
of \mbox{P/2010 A2} was discovered. We thank the Director of Space
Telescope Science Institute for allocations of Discretionary Time
used to obtain the results presented here.

\noindent \textbf{Author Contributions}

D.~J. identified \mbox{P/2010 A2} as an object of special interest,
secured HST observing time and lead the effort behind the paper.
H.~W. was responsible for the execution of the observations and
assisted with data reduction. M.~M. processed the raw images and was
responsible for the removal of cosmic rays and other artifacts.
J.~A. computed the dynamical models. M.~D. checked the work and
critiqued the proposals and paper.

\noindent \textbf{Financial Conflicts of Interest}

The authors have no competing financial interests. Correspondence
and requests for materials should be addressed to D.~J.
(jewitt@ucla.edu).

\clearpage

\begin{figure}
\centering
\includegraphics[width=\linewidth]{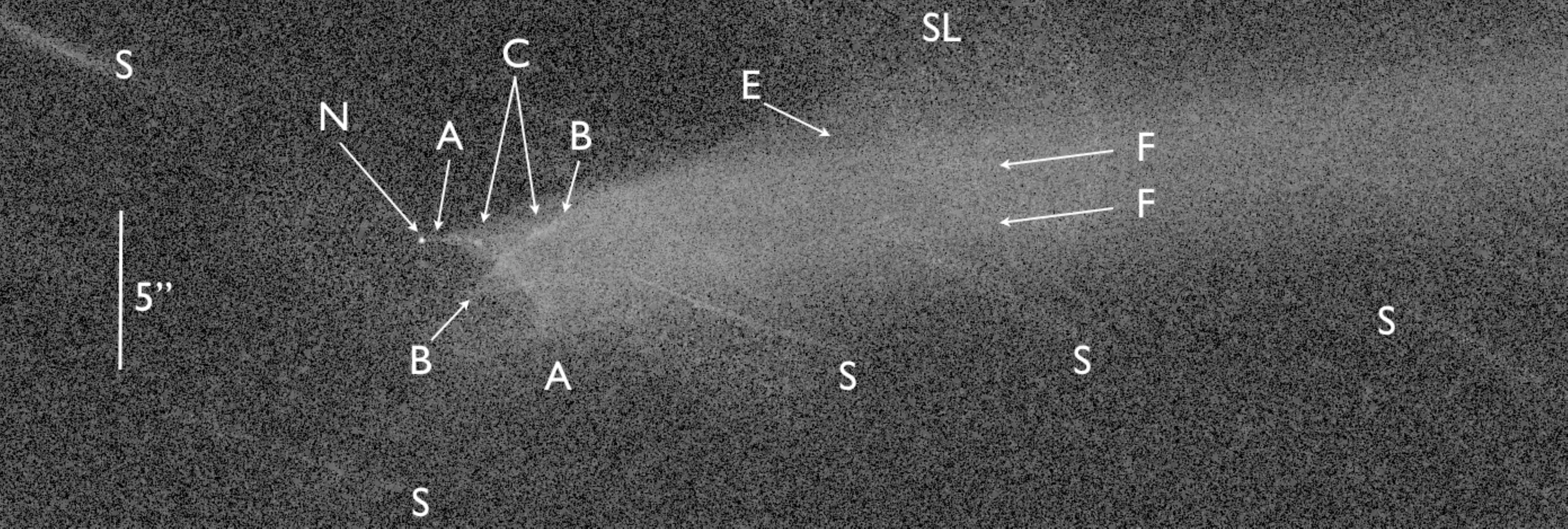}
\caption{Key to the major features in \mbox{P/2010 A2} on UT 2010
Jan 25. The principal nucleus, N, leads an arcuate dust feature, AA.
A second arcuate feature, BB, crosses AA at a large angle. Objects
at C are distinct but diffuse features detected at more than one
epoch. Particles emitted along AA and BB define the width of the
main dust tail. A separate and very diffuse dust structure, E,
extends beyond the boundaries of the tail. Linear dust streaks
(striae) are visible embedded within the tail at F. Their narrowness
shows that they emanate from discrete sources within the AA, BB arcs
with negligible initial velocity. Interfering field stars are marked
S, while SL is a band of internally scattered instrumental light
which could not be removed by image processing.}
\end{figure}

\begin{figure}
\centering
\includegraphics[width=0.9\linewidth]{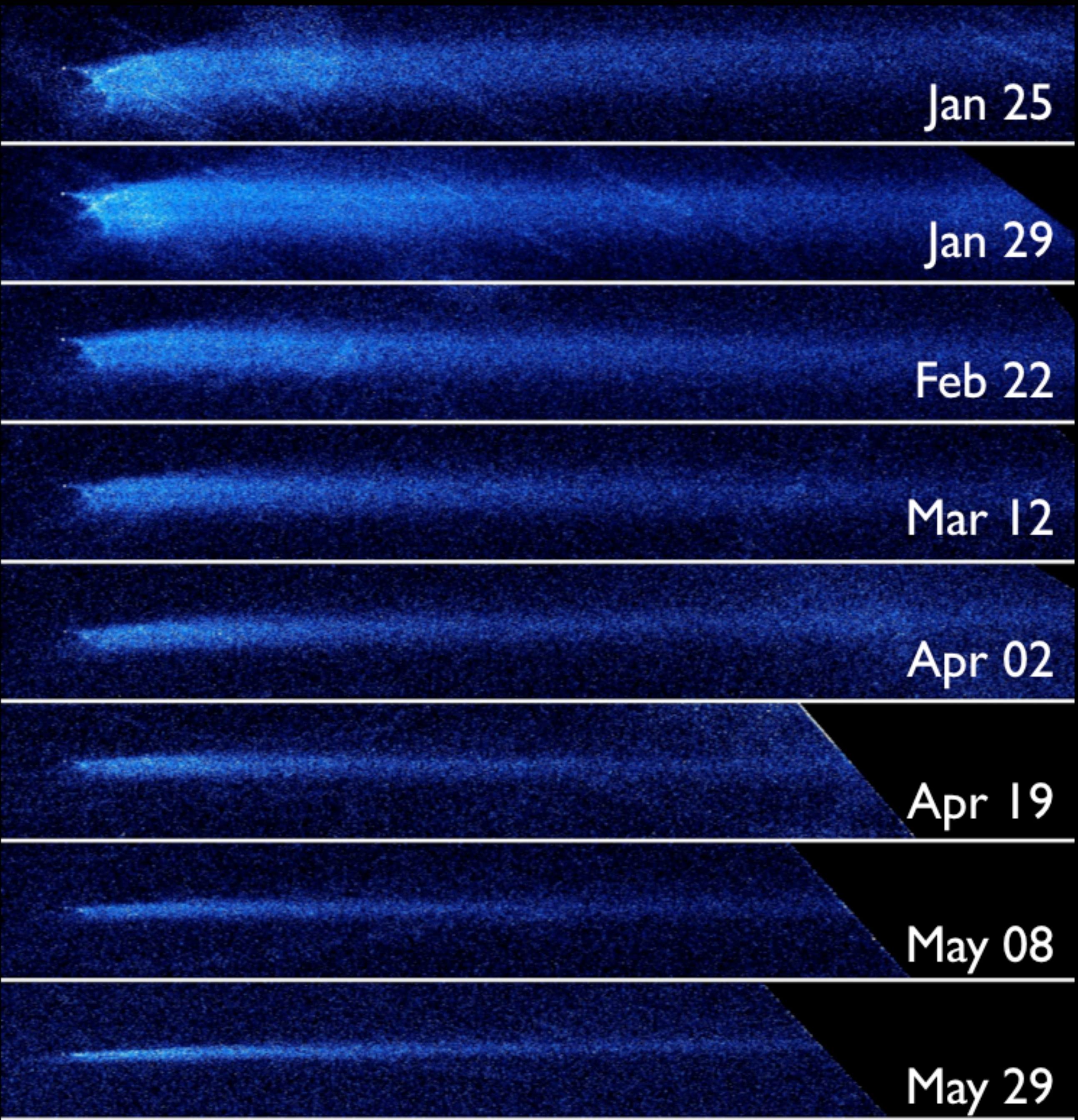}
\caption{HST images of \mbox{P/2010 A2} at the eight indicated
epochs. Images in each panel have been rotated so that the tail lies
approximately horizontally. The images, from Wide Field Camera~3,
have 0.04 arcsecond pixels and are combinations of images with total
integration times of about 2600 seconds through the F606W filter.
Each panel subtends 10 arcseconds in height. Numerous cosmic rays
and trailed background objects have been removed from the data.
Residual streaks in some panels (e.g. diagonal streaks on Jan 25 and
29) are due to the incomplete removal of trailed background stars
and galaxies.}
\end{figure}

\begin{FPfigure}
\centering
\hspace{1.65cm}\includegraphics[width=0.8\linewidth]{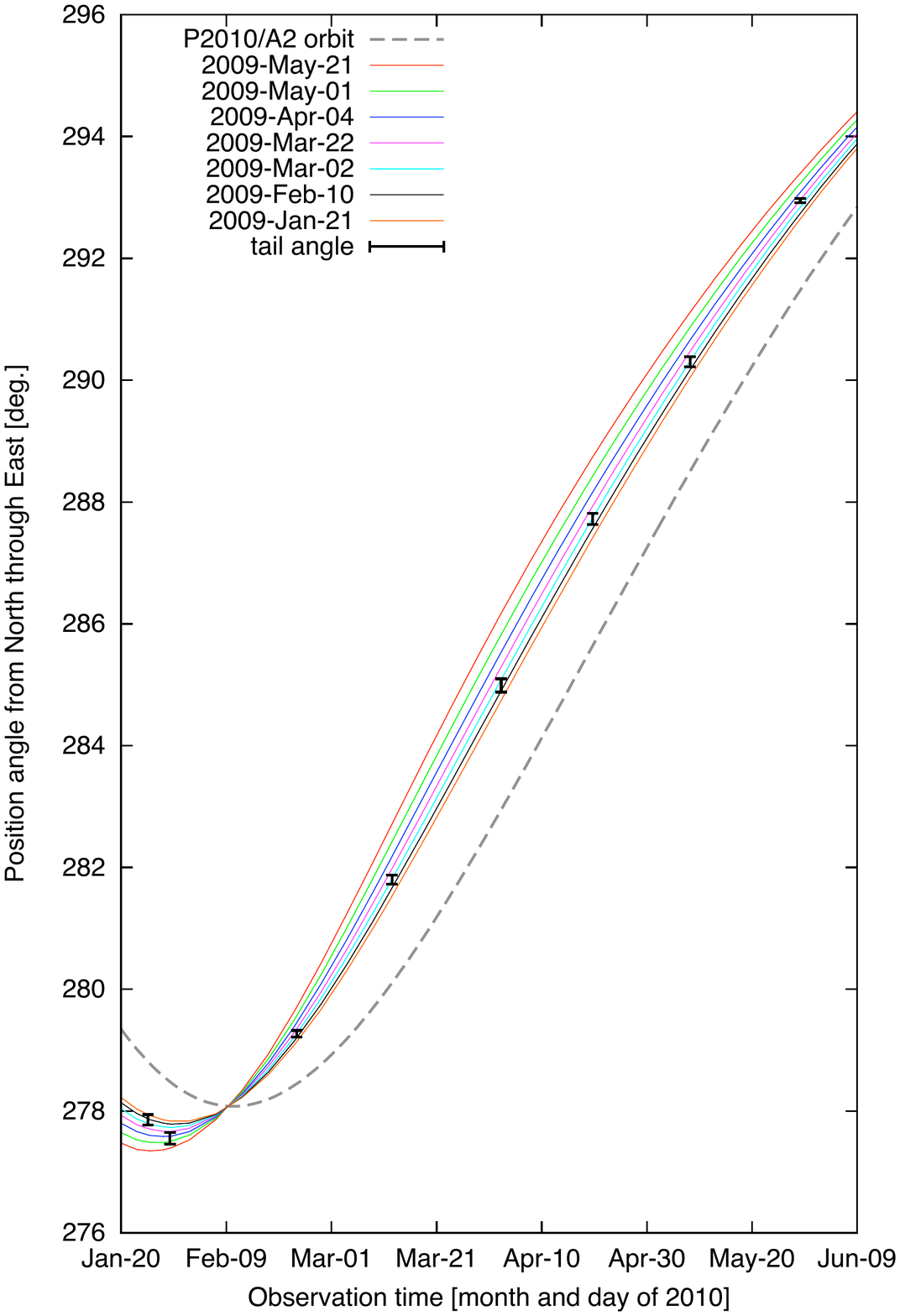}
\caption{Position angle of the tail as a function of time showing
changes caused by the viewing geometry. Measured position angles of
the tail (black symbols) are shown with error bars denoting 1
standard deviation. Calculated position angles of different
synchrones (color-coded curves) as functions of the epoch of
observation. The position angle of the projected orbit is shown in
grey. To measure the difference between the position angles of the
tail and of the projected orbit, we rotated the images such as to
align the x-axis with the projected orbit. At constant intervals, we
obtained profiles perpendicular to the orbit by averaging over 200
pixels parallel and 10 pixels perpendicular to the orbit. To each
profile we fitted a Gaussian function. We then fitted a linear
function to the peak of the Gaussian versus the distance from the
nucleus. The slope and root-mean-square of the slope give us the
position angle of the tail and the corresponding error bars. The
coloured curves indicate the position angles of specific synchrones,
i.e. dust emitted at a specific date (see label) with zero relative
velocity. Simulations demonstrate that dust emitted at a given time
with zero speed is seen in projection along a straight line starting
from the nucleus and with the distance to the nucleus proportional
to the radiation pressure coefficient, $\beta$, with larger
particles (with smaller beta) closer to the nucleus for a given
release time. For a given observation date, the position angle of
the synchrones is a unique function of the time of emission. The
coloured lines show the change of the synchrone position angles with
time, primarily due to the changing viewing geometry. In particular,
all synchrones were projected to the south of the orbit before the
Earth crossed the orbital plane of the comet on 2010 February~9, and
to the north afterwards. The measured position angles of the tail
are best matched by the 2009 March~2 synchrone and inconsistent with
synchrones more than a few weeks before or after that date.}
\end{FPfigure}


\clearpage

\begin{thebibliography}{}
\bibitem[Bottke et al.(1994)]{1994Icar..107..255B} 1. Bottke, W.~F., Nolan, M.~C., Greenberg, R., \& Kolvoord, R.~A. Velocity distributions among colliding asteroids. Icarus, 107, 255-268 (1994).
\bibitem[Holsapple(2007)]{2007Icar..187..500H} 2. Holsapple, K.~A. Spin limits of solar system bodies: From the small fast-rotators to 2003 EL61. Icarus, 187, 500-509 (2007).
\bibitem[Nesvorn{\'y} et al.(2008)]{2008ApJ...679L.143N} 3. Nesvorn{\'y}, D., Bottke, W.~F., Vokrouhlick{\'y}, D., Sykes, M., Lien, D.~J., \& Stansberry, J. Origin of the Near-Ecliptic Circumsolar Dust Band. Astrophys. J. Letters, 679, L143-146 (2008).
\bibitem[Wyatt(2008)]{2008ARA&A..46..339W} 4. Wyatt, M.~C. Evolution of Debris Disks. Annual Review of Astronomy and Astrophysics, 46, 339-383 (2008).
\bibitem[Birtwhistle et al.(2010)]{2010CBET.2114....1B} 5. Birtwhistle, P., Ryan, W.~H., Sato, H., Beshore, E.~C., \& Kadota, K. Comet \mbox{P/2010 A2} (LINEAR). Central Bureau Electronic Telegrams, 2114, 1 (2010).
\bibitem[Hsieh \& Jewitt(2006)]{2006Sci...312..561H} 6. Hsieh, H.~H., \& Jewitt, D. A population of comets in the main asteroid belt. Science, 312, 561-563 (2006).
\bibitem[Jewitt et al.(2010)]{2010IAUC.9109....3J} 7. Jewitt, D., Annis, J., \& Soares-Santos, M. Comet \mbox{P/2010 A2} (LINEAR). \iaucirc, 9109, 3 (2010).
\bibitem[Licandro et al.(2010)]{2010CBET.2134....3L} 8. Licandro, J., Tozzi, G.~P., Liimets, T., Cabrera-Lavers, A., \& Gomez, G. Comet \mbox{P/2010 A2} (LINEAR). Central Bureau Electronic Telegrams, 2134, 3 (2010).
\bibitem[Gradie \& Tedesco(1982)]{1982Sci...216.1405G} 9. Gradie, J., \& Tedesco, E. Compositional structure of the asteroid belt. Science, 216, 1405-1407 (1982).
\bibitem[Agarwal et al.(2010)]{2010arXiv1001.3010A} 10. Agarwal, J., Mueller, M., Reach, W.T., Sykes, M.V., Boehnhardt, H. \& Gruen, E. The dust trail of Comet 67P/Churyumov-Gerasimenko between 2004 and 2006. Icarus 207, 992-1012 (2010).
\bibitem[Snodgrass et al.(2010)]{this_issue} 11. Snodgrass, C. et al. Recent asteroid collision \mbox{P/2010 A2} confirmed and dated by Rosetta/OSIRIS observations. Nature, this issue (2010).
\bibitem[Rubincam(2000)]{2000Icar..148....2R} 12. Rubincam, D.~P. Radiative Spin-up and Spin-down of Small Asteroids. Icarus, 148, 2-11 (2000).
\bibitem[Taylor et al.(2007)]{2007Sci...316..274T} 13. Taylor, P.~A., et al. Spin rate of asteroid (54509) 2000 PH5 increasing due to the YORP effect. Science, 316, 274-277 (2007).
\bibitem[Benz \& Asphaug(1999)]{1999Icar..142....5B} 14. Benz, W., \& Asphaug, E. Catastrophic Disruptions Revisited. Icarus, 142, 5-20 (1999).
\bibitem[Michikami et al.(2007)]{2007P&SS...55...70M} 15. Michikami, T., Moriguchi, K., Hasegawa, S., \& Fujiwara, A. Ejecta velocity distribution for impact cratering experiments on porous and low strength targets. Planetary and Space Sciences, 55, 70-88 (2007).
\bibitem[Jutzi et al.(2010)]{2010Icar..207...54J} 16. Jutzi, M., Michel, P., Benz, W., \& Richardson, D.~C. Fragment properties at the catastrophic disruption threshold: The effect of the parent body's internal structure. Icarus, 207, 54-65 (2010).
\bibitem[Bottke et al.(2005)]{2005Icar..179...63B} 17. Bottke, W.~F., Durda, D.~D., Nesvorn{\'y}, D., Jedicke, R., Morbidelli, A., Vokrouhlick{\'y}, D., \& Levison, H.~F. Linking the collisional history of the main asteroid belt to its dynamical excitation and depletion. Icarus, 179, 63-94 (2005).
\bibitem[Leinert et al.(1983)]{1983A&A...118..345L} 18. Leinert, C., Roser, S., \& Buitrago, J. How to maintain the spatial distribution of interplanetary dust. Astron. Astrophys., 118, 345-357 (1983).
\bibitem[Nesvorn{\'y} et al.(2010)]{2010ApJ...713..816N} 19. Nesvorn{\'y}, D., Jenniskens, P., Levison, H.~F., Bottke, W.~F., Vokrouhlick{\'y}, D., \& Gounelle, M. Cometary origin of the zodiacal cloud and carbonaceous micrometeorites: implications for hot debris disks. Astrophys. J., 713, 816-836 (2010).
\end{thebibliography}
\end{document}